\documentstyle[aps]{revtex}
\begin{document}
\voffset 0.8truecm
\title{
A universal cloner allowing the
input to be arbitrary states in
symmetric subspace
}

\author{
Heng Fan$^a$, Keiji Matsumoto$^a$, Xiang-Bin Wang$^a$,
Hiroshi Imai$^a$, and Miki Wadati$^b$}
\address{
$^a$Quantum computing and information project,
ERATO, \\
Japan Science and Technology Corporation,\\
Daini Hongo White Bldg.201, Hongo 5-28-3, Bunkyo-ku, Tokyo 133-0033, Japan.\\
$^b$Department of Physics, Graduate School of Science,\\
University of Tokyo, Hongo 7-3-1, Bunkyo-ku, Tokyo 113-0033, Japan.
}
\maketitle
                                                        
\begin{abstract}
A generalized universal quantum cloning machine is
proposed which allows the input to be arbitrary states
in symmetric subspace.
And it reduces to the universal quantum cloning machine (UQCM)
if the input are identical pure states.
The generalized cloner is optimal
in the sense we compare the input and output
reduced density operators at a single qubit.
The result for
qubits is extended to arbitrary-dimensional states.
\end{abstract}
       

Quantum information theory\cite{BS} has been attracting
a great deal of interests.
The no-cloning theorem describes one of the most fundamental
nonclassical properties of quantum systems.
There are many applications of this theorem, for example,
to quantum cryptography.
It states that
an unknow quantum state cannot be cloned exactly\cite{WZ,D},
but only approximately.
And the no-cloning theorem for pure states is extended to
other cases\cite{BCFJS,M,KI}. However, the no-cloning theorem
does not forbid imperfect cloning. We are interested
to know how we can clone quantum states
as good as possible.
And much effort has been put into developing optimal
cloning processes.
Bu\v{z}ek and Hillery \cite{BH}
proposed a 1 to 2 UQCM which produecs two indentical copies
whose quality is independent of the input qubit (two-level
system). This UQCM was later proved to be optimal if
the measure of quality is the fidelity between the input
and the ouput\cite{BDEFMS}. Gisin and Massar \cite{GM}
considered a general case in which $M$ identical
copies are generated from $N$ ($M\geq N$) identical qubits,
and proved that the fidelity of the copies is optimal.
The connection
between optimal quantum cloning and the optimal state
estimation was introduced in \cite{BEM} and
a tight upper bound for the
fidelity of $N$ to $M$ UQCM was obtained. The
complete positive (CP) map of the optimal
$N$ to $M$ cloning transformation and the optimal
fidelity for arbitrary dimensional system were derived
and studied extensively by Werner and
Keyl and Werner in \cite{W,KW}.  Besides the
pure input states,
Bu\v{z}ek and Hillery also formulated the 1 to 2 UQCM for
d-dimensional Hilbert space, and studied the cloning
of impure states and the cloning of entangled state of
two qubits\cite{BH1}. The optimal quantum cloning via
stimulated emission was proposed in \cite{SWZ}.
The copying network for UQCM was presented in \cite{BBHB}.
And universal cloning of continuous quantum variables
was studied in \cite{CIR,BCILM}.
Other quality measure of cloning based on distinguishability of
states is studied in \cite{NG}.

In this paper, we study the quantum cloning of arbitrary states
in symmetric subspace.
We first restrict our discussions to the 2-level
($|\uparrow \rangle,|\downarrow \rangle$) quantum system.
The input is an arbitrary density operator of M qubits in
symmetric subspace,
\begin{eqnarray}
\rho ^{in}(M)=\sum _{j,j'=0}^Mx_{jj'}|(M-j)\uparrow , j\downarrow 
\rangle \langle j'\downarrow , (M-j')\uparrow |.
\label{arbinput}
\end{eqnarray}
Here $|(M-j)\Psi ,j\Psi ^{\perp}\rangle $ is
the symmetric and normalized state with $M-j$ qubits in the
state $\Psi $ and $j$ in the orthonormal state $\Psi ^{\perp}$
which is invariant under all permutations.
We take $|\Psi \rangle =|\uparrow \rangle ,
|\Psi ^{\perp}\rangle =|\downarrow \rangle $,
$x_{jj'}$ is an arbitrary matrix and
we let $\sum _{j=0}^{M}x_{jj}=1$ which is the trace
condition for density operators.
We remark that $M$ identical pure input states
$|M\Psi \rangle\equiv |\Psi \rangle ^{\otimes M}$
is a special case of (\ref{arbinput}).
The reduced density operators of (\ref{arbinput})
at each qubit are the same and take the form
\begin{eqnarray}
\rho ^{in}_{red.}(M)
&=&|\uparrow \rangle \langle \uparrow |
\sum _{j=0}^Mx_{jj}\frac {M-j}{M}
+|\uparrow \rangle \langle \downarrow |
\sum _{j=0}^{M-1}x_{jj+1}\frac {\sqrt{(M-j)(j+1)}}{M}
\nonumber \\
&&+|\downarrow \rangle \langle \downarrow |
\sum _{j=0}^Mx_{jj}\frac {j}{M}
+|\downarrow \rangle \langle \uparrow | 
\sum _{j=0}^{M-1}x_{j+1j}\frac {\sqrt{(M-j)(j+1)}}{M}.
\label{redin}
\end{eqnarray}
The goal of this paper is to find the optimal cloning transformation
with input (\ref{arbinput}) and
output $\rho ^{out}(M,L)$ in $L$ qubits, so that the fidelity
between
$\rho ^{in}_{red.}(M)$ in (\ref{redin}) and
the output reduced density operator at each qubit
$\rho ^{out}_{red.}(M,L)$ can achieve the upper bound.
We call the cloning transformation with (\ref{arbinput}) as input
a generalized UQCM (g-UQCM) in this paper to distinguish it
from UQCM which takes identical pure states as input \cite{BH,GM}.
The relation between input and output reduced density operators
can be written in a scaling form  
\begin{eqnarray}
\rho ^{out}_{red.}
=\eta (M,L)\rho ^{in}_{red.}+{1\over 2}(1-\eta (M,L)),
\end{eqnarray}
where $\eta (M,L)$ is the shrinking factor of Bloch vector
characterizing the operation of universal quantum cloning transformation.
The optimal g-UQCM refer to maximal $\eta (M,L)$.
By identifying the optimal
fidelity of $M$ to $\infty$ cloning 
with optimal fidelity of quantum state estimation
for $M$ identical unknown pure states \cite{MP,DBE},
Bru\ss , Ekert and Macchiavello \cite{BEM} obtained the
tight upper bound of the shrinking factor, 
$\eta (M,L)=\frac {M(L+2)}{L(M+2)}$.

With $M$ identical pure qubits $|M\Psi \rangle $ as input,
Bu\v{z}ek-Hillery ($1\rightarrow 2$) and Gisin-Massar ($M\rightarrow L$)
UQCM which achieve the optimal shrinking factor $\eta (M,L)$
have already been proposed\cite{BH,GM}.
It is explicit that the input $|M\Psi \rangle $ belongs
to symmetric subspace.
Because we use the fidelity between input and output reduced
density operators at a single qubit to define the quality of cloning
for both UQCM and g-UQCM. The g-UQCM will reduce
to UQCM if the input are $M$ identical pure states $|M\Psi \rangle $.
We can also study the concatenation of two quantum cloners\cite{BEM}.
The first one is the UQCM which
acts on $N$ identical pure qubits $|N\Psi \rangle $
and produces $M$ copies, and the
second cloner uses the output of first cloner as input and generates
$L$ copies.
The output of a UQCM which is generally an entangled and/or
mixed state
belongs to symmetric subspace. Thus the second cloner can be
formulated by a g-UQCM.

We propose the unitary
cloning transformation of the g-UQCM as follows
\begin{eqnarray}
U(M,L)
|(M-j)\Psi ,j\Psi ^{\perp}\rangle \otimes R
=\sum _{k=0}^{L-M}
\alpha _{jk}(M,L)|(L-j-k)\Psi ,(j+k)\Psi ^{\perp}\rangle \otimes R_k,
\label{second}
\end{eqnarray}
where
\begin{eqnarray}                               
\alpha _{jk}(M,L)=\sqrt{\frac {(L-M)!(M+1)!(L-j-k)!(j+k)!}
{(L+1)!(L-M-k)!(M-j)!j!k!}},\nonumber \\
j=0,\cdots, M;~~k=0, \cdots, L-M,
\label{second1}
\end{eqnarray}
$R$ denotes the initial state of the UQCM and
$M-N$ blank copies, $R_j$ are the orthonormalized internal
states of the UQCM (ancilla states),
In case $j=0$, it
reduces to the original UQCM with $M$ identical pure input states
and $L$ copies\cite{GM}.
This g-UQCM allows the input to be
not only identical pure states but also mixed and/or entangled
states in symmetric subspace. 
We now show that this g-UQCM is still optimal in the sense
that the shrinking factor between input and output reduced
density operators at each qubit achieves the upper bound.
Substituting $|\Psi \rangle =|\uparrow \rangle ,
|\Psi ^{\perp}\rangle =|\downarrow \rangle $ into (\ref{second}),
and applying this cloning transformation on 
the input density operator (\ref{arbinput}),
$U(M,L)\rho ^{in}(M)U^{\dagger }(M,L)$,
taking trace over ancilla states, we can obtain
the output density operator with $L$ qubits. The reduced
output density operator of each qubit is derived as
\begin{eqnarray}
\rho ^{out}_{red.}(M,L)
&=&|\uparrow \rangle \langle \uparrow |
\sum _{j=0}^M\sum _{k=0}^{L-M}
x_{jj}\alpha _{jk}^2(M,L)\frac {L-j-k}{L}
+|\downarrow \rangle \langle \downarrow |
\sum _{j=0}^M\sum _{k=0}^{L-M}x_{jj}\alpha _{jk}^2(M,L)
\frac {j+k}{L}
\nonumber \\
&&+|\uparrow \rangle \langle \downarrow |
\sum _{j=0}^{M-1}\sum _{k=0}^{L-M}
x_{jj+1}\alpha _{jk}(M,L)\alpha _{j+1k}(M,L)
\frac {\sqrt{(L-j-k)(j+k+1)}}{L}
\nonumber \\
&&+|\downarrow \rangle \langle \uparrow | 
\sum _{j=0}^{M-1}\sum _{k=0}^{L-M}
x_{j+1j}\alpha _{jk}(M,L)\alpha _{j+1k}(M,L)
\frac {\sqrt{(L-j-k)(j+k+1)}}{L}.
\label{redout}
\end{eqnarray}
Comparing (\ref{redout})
with the reduced input density operator $\rho ^{in}_{red.}(M)$ in
(\ref{redin}) at
each qubit of input state (\ref{arbinput}), and after some calculations,
we have
\begin{eqnarray}
\rho ^{out}_{red.}(M,L)=\frac {M(L+2)}{L(M+2)}\rho ^{in}_{red.}(M)
+\frac {L-M}{L(M+2)}\cdot 1.
\label{scaling}
\end{eqnarray}
In the calculations, only trace condition of the input
density operator is used, the positivity
condition of the density operator is not used.
That means we even do not need (\ref{arbinput}) 
as a density operator, but the scaling form of
cloning (\ref{scaling}) is still hold.
Thus we see that the shrinking factor characterizing the
g-UQCM (\ref{second}) achieves the upper bound and
is independent from the arbitrary input density operators
(\ref{arbinput}) in symmetric subspace. The unitary cloning
transformation (\ref{second},\ref{second1}) is a universal
and optimal cloner which allows the input to be arbitrary
states in symmetric subspace.

As an example, we study the concatenation of a UQCM and a g-UQCM. 
Taking $|N\Psi \rangle $ as input, using cloning
transformation (\ref{second}),
tracing over the ancilla states $R_j$, we can
obtain the output density operator of $M$ copies as
\begin{eqnarray}
\rho ^{out}(N,M)=\sum _{j=0}^{M-N}\alpha _{0j}^2(N,M)
|(M-j)\Psi ,j\Psi ^{\perp }\rangle \langle j\Psi ^{\perp},
(M-j)\Psi | .
\label{nmout}
\end{eqnarray}
We remark that (\ref{nmout}) is the output density operator
of a UQCM proposed by Gisin and Massar \cite{GM}.
We now concatenate a g-UQCM to the
$N$ to $M$ UQCM with (\ref{nmout}) as input
and produce $L$ copies.
Using the cloning transformation (\ref{second},\ref{second1}),
the output density operator of the g-UQCM takes the form
\begin{eqnarray}
\rho ^{out}(N,M,L)&=&\sum _{j=0}^{M-N}\sum _{k=0}^{L-M}
\alpha _{0j}^2(N,M)\alpha ^2_{jk}(M,L)|(L-j-k)\Psi ,(j+k)\Psi ^{\perp}\rangle
\langle (j+k)\Psi ^{\perp},(L-j-k)\Psi |
\nonumber \\
&=&\sum _{p=0}^{L-N}\alpha _{0p}^2(N,L)
|(L-p)\Psi ,p\Psi ^{\perp }\rangle \langle p\Psi ^{\perp},
(L-p)\Psi | ,
\end{eqnarray}
where we have used a simple relation
which can be derived
from $(x+y)^{M-N}(x+y)^{L-M}=(x+y)^{L-N}$ to obtain the last
equation. We can find the output density
operator of the sequence of the concatenated cloners is
the same as the output density operator of $N$ to $L$ UQCM,
$\rho ^{out}(N,M,L)=\rho ^{out}(N,L)$.
We already know that two UQCM are optimal, it is straightforward that
the g-UQCM (\ref{second},\ref{second1}) is optimal,
otherwise it would lead to a contradiction.
We have shown here another method to prove
the optimum of the g-UQCM 
in the case when input are identical pure states
or the output density operator produced by a UQCM.

Next, we study the d-level quantum system. Quantum cloning
with $N$ identical pure input states and $M$ copies
in arbitrary d-dimensional Hilbert spaces is
formulated by CP map in \cite{W,KW}, and the
optimal fidelity is given as,
$F(d: N,M)=\frac {N(M+d)+M-N}{M(N+d)}$.
With the result of the optimal fidelity for d-level
quantum cloning, the optimal fidelity of
state estimation for finite and identical d-level
quantum states can be obtained \cite{BM}.
Similar to 2-level (qubit) case, the density matrix of
d-level state can
be expressed by generalized Bloch vector
${\vec s}=(s_1, \cdots, s_{d^2-1})$ and the
generators $\tau _i, i=1, \cdots , d^2-1$ of the group $SU(d)$,
$\rho ={1\over d}+{1\over 2}\sum _{i=1}^{d^2-1}s_i\tau _i$,
where the generators of $SU(d)$
is defined as ${\rm Tr}\tau _i=0, ~{\rm Tr}(\tau _i\tau _j)=2\delta _{ij}$.
With $N$ identical pure states as input, the
reduced output density operator at each d-level state
of $N$ to $M$ UQCM takes the form
$\rho ^{out}={1\over d}+{1\over 2}
\eta (d:N,M)\sum _{i=1}^{d^2-1}s_i\tau _i$.
Corresponding to optimal
fidelity, the upper bound of the shrinking factor for both
UQCM and g-UQCM is
$\eta (d:N,M)=\frac {N(M+d)}{M(N+d)}$.

1 to 2 unitary cloning transformation of d-level system was formulated
in \cite{BH1}. 1 to M and a special case of N to M cloning
transformation was given in \cite{AF},
and the general unitary $N$ to $M$ UQCM was given in \cite{FKW}
where the form is different from this paper
(in \cite{W,KW} the
CP map of the general cloning transformation was derived).
Similar to Gisin-Massar cloner, we present here
the UQCM for d-level system.
Let $|\Psi \rangle $ be
an arbitrary state in d-level system,
$|\Psi ^{\perp}_1\rangle,
\cdots ,|\Psi ^{\perp}_{d-1}\rangle $ be othonormal
states. The d-level $N$ to $M$ UQCM takes the following form,
\begin{eqnarray}
U(d:N,M)|N\Psi \rangle\otimes R&=&
\sum _{\vec j}^{M-N}\alpha _{\vec j}(N,M)
|(N+j_0)\Psi , j_1\Psi ^{\perp}_1,
\cdots ,j_{d-1}\Psi ^{\perp}_{d-1}\rangle \otimes R_{\vec j},
\nonumber \\
\alpha _{\vec j}(N,M)&=&\sqrt{\frac {(M-N)!(N+d-1)}{(M+d-1)!}}
\sqrt{\frac {(N+j_0)!}{N!j_0!}},
\label{dclone}
\end{eqnarray}
where $\vec {j}=(j_0, j_1, \cdots , j_{d-1})$,
state $|(N+j_0)\Psi , j_1\Psi ^{\perp}_1,
\cdots ,j_{d-1}\Psi ^{\perp}_{d-1}\rangle $ is a completely symmetric
and normalized state with $N+j_0$ states in $\Psi $,
$j_i$ states in $\Psi ^{\perp}_i, i=1, \cdots ,d-1$,
and summation $\sum _{\vec j}^{M-N}$ means sum over all variables under the
condition $\sum _{i=0}^{d-1}j_i=M-N$.
$R_{\vec k}$ are orthonormal internal states of the cloner,
$\langle R_{\vec k}|R_{\vec k'}\rangle =\delta _{\vec {k}\vec {k'}}$.
We next prove that this cloning
transformation is the optimal UQCM. Since the
optimal fidelity is already available\cite{W}, we just
need to prove the fidelity of the cloning transformation (\ref{dclone})
achieves this upper bound.
As the qubits case\cite{GM}, the fidelity of d-level UQCM can be
calculated as
\begin{eqnarray}
F(d:N,M)=\sum _{\vec j}^{M-N}\alpha _{\vec j}^2(N,M)\frac {(N+j_0)}{M}
=\frac {N(M+d)+M-N}{M(N+d)},
\label{dfidelity}
\end{eqnarray}
where $\alpha _{\vec j}^2(N,M)$ is the probability of state
$|(N+j_0)\Psi , j_1\Psi ^{\perp}_1,
\cdots ,j_{d-1}\Psi ^{\perp}_{d-1}\rangle \otimes R_{\vec j}$ in
the output, and $\frac {(N+j_0)}{M}$ is the ratio of the
number of ways to choose $(N+j_0-1)\Psi ,j_1\Psi _1^{\perp },
\cdots ,j_{d-1}\Psi ^{\perp}_{d-1}$ among $M-1$ d-level states
over the number of ways to choose
$(N+j_0)\Psi ,j_1\Psi _1^{\perp },
\cdots ,j_{d-1}\Psi ^{\perp}_{d-1}$ among $M$ d-level states.
The fidelity (\ref{dfidelity}) of d-level UQCM (\ref{dclone})
is optimal, thus (\ref{dclone}) is the optimal UQCM
and the shrinking factor achieves its upper bound.
Tracing out the ancilla, we have the output density operator
\begin{eqnarray}
\rho ^{out}(d:N,M)=
\sum _{\vec j}^{M-N}\alpha _{\vec j}^2(N,M)
|(N+j_0)\Psi , j_1\Psi ^{\perp}_1,
\cdots ,j_{d-1}\Psi ^{\perp}_{d-1}\rangle
\langle j_{d-1}\Psi ^{\perp}_{d-1}, \cdots, j_1\Psi ^{\perp}_1,
(N+j_0)\Psi |
\label{dout}
\end{eqnarray}

We finally propose a g-UQCM which allows
arbitrary states with $M$ d-level states
belonging to the symmetric subspace
as input, and produces $L$ copies. The cloning transformation
takes the form,
\begin{eqnarray}
&&U(d:M,L)|j_0\Psi ,j_1\Psi _1^{\perp} ,\cdots ,
j_{d-1}\Psi _{d-1}^{\perp }\rangle \otimes R
\nonumber \\
&&=\sum _{\vec k}^{L-M}\alpha _{\vec {j}\vec {k}}(M,L)
|(j_0+k_0)\Psi ,(j_1+k_1)\Psi _1^{\perp} ,\cdots ,
(j_{d-1}+k_{d-1})\Psi _{d-1}^{\perp }\rangle \otimes R_{\vec k},
\label{dgeneral} \\
&&\alpha _{\vec{j},\vec {k}}(M,L)=
\sqrt{\frac {(L-M)!(M+d-1)!}{(L+d-1)!}}
\sqrt{\prod _{i=0}^{d-1}\frac {(j_i+k_i)!}{j_i!k_i!}}.
\end{eqnarray}
where $\sum _{i=0}^{d-1}j_i=M, \sum _{i=0}^{d-1}k_i=L-M$ are assumed.
The optimum of this g-UQCM for d-level system can
be proved by a similar method as for 2-level system.
Using the output density operator (\ref{dout}) as input,
applying the cloning transformation (\ref{dgeneral}),
one can prove the output of $\rho ^{out}(d:N,M,L)$
is the same as the output $\rho ^{out}(d:N,L)$ of one $N$ to $L$ UQCM.
In the calculations, a relation derived from an
identity $(\sum _{i=0}^{d-1}x_i)^{M-N}(\sum _{i=0}^{d-1}x_i)^{L-M}
=(\sum _{i=0}^{d-1}x_i)^{L-N}$ is useful.
We already know that the $N$ to $M$ UQCM and $N$ to $L$ UQCM
are optimal. The g-UQCM which
use the output density operator (\ref{dout}) as input
and generates $L$ copies is thus optimal.
The g-UQCM (\ref{dgeneral}) allows the input
to be mixed and/or entangled states supported in
symmetric subspace, and the shrinking
factor achieves the upper bound
$\eta (d:M,L)=\frac {M(L+d)}{L(M+d)}$.
We remark that the dimension of the internal state of the
cloner (ancilla) is $\frac {(L-M+d-1)!}{(L-M)!(d-1)!}$ which
is useful in POVM (positive operator valued measurement),
see for example \cite{PW,FP,S}.
Note the ancilla states $R_{\vec {k}}$
should be expressed more precisely as $R_{\vec{k}}(\Psi )$ and
can be realized in symmetric subspace
$R_{\vec{k}}(\Psi )=|k_0\Psi , k_1\Psi ^{\perp}_1, \cdots ,
k_{d-1}\Psi ^{\perp }_{d-1}\rangle $.

Suppose d-level quantum system is spanned by the
orthonormal basis $|i\rangle, i=0, \cdots ,d-1$, an arbitrary
pure state is written as the form
$|\Phi \rangle =\sum _{i=0}^{d-1}c_i|i\rangle $ with
$\sum _{i=0}^{d-1}|c_i|^2=1$. Then any $M$ d-level states
in symmetric subspace can be expressed as
$|\vec {j}\rangle $, where $j_i$ states are
in $|i\rangle ,i=0, \cdots, d-1$, and $\sum _{i=0}^{d-1}j_i=M$.
We take a special case;
let $|\Psi \rangle =|0\rangle ,|\Psi ^{\perp }_i\rangle =|i\rangle ,
i=1, \cdots ,d-1$. The $M$ to $L$ quantum cloning transformation
(\ref{dgeneral}) can be rewritten as,
\begin{eqnarray}
&&U(d:M,L)|\vec{j}\rangle \otimes R
=\sum _{\vec k}^{L-M}\alpha _{\vec {j}\vec {k}}(M,L)
|\vec {j}+\vec {k} \rangle \otimes R_{\vec k},
\label{another}
\end{eqnarray}
where we still denote the internal states of the cloner by 
$R_{\vec k}$ in this special case for convenience. These results
coincide with our previous formulae\cite{FKW} where only identical
pure input states are studied. In this paper, we study the
input to be arbitrary states in symmetric subspace.
Because $|\vec {j}\rangle , \sum _{i=0}^{d-1}j_i=M$, can
be the orthonormal basis for $M$ states d-level system
in symmetric subspace,
this cloning transformation (\ref{another}) is another
independent and complete set of cloning transformation
equivalent to (\ref{dgeneral}).
This cloning transformation is optimal and achieves
the upper bound of the shrinking factor.
When the input is identical pure states, it is
the UQCM, and two or more cloners can be concatenated together.
As the qubits case, the g-UQCM can be
used as a concatenated cloner, and the
input can be arbitrary states in symmetric subspace.
The input state consisting of $M$ d-dimensional states
in symmetric subspace is written as
$\rho ^{in}(d:M)=\sum _{\vec {j}\vec {j'}}^Mx_{\vec {j}\vec{j'}}
|\vec {j}\rangle \langle \vec {j'}|$.
The dimension of matrix $x_{\vec {j}\vec {j'}}$ is
$\frac {(M+d-1)!}{M!(d-1)!}$, and we let
$\sum _{\vec {j}}^Mx_{\vec {j}\vec {j}}=1$.
Using the cloning transformation (\ref{another}),
the reduced density operator of output can still have
an optimal shrinking factor  
$\eta (d:M,L)=\frac {M(L+d)}{L(M+d)}$
compared with the reduced density operator of input,
$\rho ^{out}_{red.}(d:M,L)=\eta (d:M,L)
\rho ^{in}_{red}(M)+{1\over d}(1-\eta (d:M,L))$.
We should note, 
for 1 to 2 cloning, the universal cloning of arbitrary impure
states was studied by Bu\v{z}ek and Hillery \cite{BH1}
corresponding to the case $M=1, L=2$ in (\ref{another}).
                                  
In conclusion, we have proposed a g-UQCM which
allows arbitrary input states belonging to symmetric
subspace. The g-UQCM is optimal. Bu\v{z}ek and
Hillery \cite{BH1} studied the optimal cloning of
two qubits entangled states by a 1 to 2, 4-level UQCM.
The goal of this paper is that
the input of cloning is not only the
pure states but also arbitrary states in symmetric subspace.
The optimum of the g-UQCM is in
the sense that the
shrinking factor (fidelity) between
input and output reduced density
operators in one d-level state achieves its upper bound. This g-UQCM
reduces to UQCM if the input is identical pure states.

We would like to thank M.Koashi,G.Weihs and T.Yamakami for useful
discussions.

\end{document}